\documentclass[preprint,number,3pd]{elsarticle}
\usepackage{amsmath}
\usepackage{amssymb}
\usepackage{graphicx}

\def\be{\begin{equation}}
\def\bsp{\begin{split}}
\def\a{\alpha}
\def\b{\beta}

\def\d{\delta}
\def\D{\Delta}
\def\e{\epsilon}

\def\m{\mu}

\def\p{\pi}
\def\r{\rho}

\def\S{\Sigma}
\def\t{\tau}
\def\f{\varphi}

\def\c{\chi}
\def\ps{\psi}

\def\O{\Omega}
\def\i{\int}

\def\bff{{\mathbf f}}
\def\by{{\mathbf y}}
\def\bx{{\mathbf x}}
\def\bX{{\mathbf X}}

\def\bj{{\mathbf j}}

\def\bff{{\mathbf f}}

\def\bS{{\mathbf S}}

\def\ee#1{\label{#1}\end{equation}}
\def\esq{\end{split}}

\begin{document}
\title{Rate description of Fokker-Planck processes with time-periodic
parameters
}
\author[Au,Te]{Changho Kim}
\ead{changhokim@kaist.ac.kr}
\author[Au]{Peter Talkner \corref{cor1}}
\ead{peter.talkner@physik.uni-augsburg.de}
\author[Te]{Eok Kyun Lee}
\ead{eklee@cola.kaist.ac.kr}
\author[Au]{Peter H\"anggi}
\ead{hanggi@physik.uni-augsburg.de}
\cortext[cor1]{corresponding author}
\address[Au]{Universit\"at Augsburg, Institut f\"ur Physik,
D-86135 Augsburg, Germany}
\address[Te]{Department of Chemistry and School of Molecular Science
  (BK21),Korea Advanced Institute of Science and Technology, Daejeon
  305-701, Republic of Korea}
\date{\today}
\begin{abstract}
The large time dynamics of a periodically driven Fokker-Planck process
possessing several metastable states is investigated. At weak noise
transitions between the metastable states are rare. Their dynamics then
represent a discrete Markovian process characterized by time
dependent rates. Apart from the occupation probabilities, so-called specific
probability densities and localizing functions can be associated to each
metastable state.
Together, these three sets of functions uniquely
characterize the large time dynamics of the conditional probability
density of the original process. Exact equations of motion are
formulated for these three sets of functions and strategies are
discussed how to solve them. These methods are illustrated and their
usefulness is demonstrated by means
of the example of a bistable Brownian oscillator within a large range
of driving frequencies from the slow semiadiabatic  to the fast
driving regime.
\end{abstract}
\maketitle

\section{Introduction}\label{I}
Chemical reactions have provided ubiquitous and versatile examples of
activated transitions between two metastable states, formed by the
reactants and products. In a chemical
reaction the energy necessary for the activation most often stems from the
(classical or even quantum mechanical) thermal energy that may accumulate in a single reaction
coordinate and finally enable a transition from reactants to products
\cite{P1986,PGH,HTB, PT}. In contrast to these thermally assisted escape processes  other
additional
sources of energy may externally be provided for example by driving
a system with metastable states by periodic forces.
Such periodically driven stochastic systems present a particular class of
nonequilibrium processes that exhibit a broad variety of fascinating
effects \cite{JH90,JH,J} such as stochastic resonance \cite{SR}, directed
transport of Brownian particles in ratchet
type periodic potentials \cite{AH,BM,HM}
or other anomalous transport properties
as for example negative mobility \cite{NM}.
Apart from an external periodic driving, these systems typically are subject to
nonlinear dynamical laws and additionally experience
fluctuating forces describing the random impact of the environment of
the considered system \cite{Z}. Without the fluctuating forces the presence of
nonlinearities often renders these systems multistable, i.e.\ such
systems may approach different attractors \cite{P1978}, depending on
their initial 
states. In combination with weak fluctuating forces these attractors
become metastable states, which means that the system will be found
most of the time in
or close to one of these states while transitions between these states
present rare events.

Each of the principal constituents of the dynamics of a periodically
driven nonlinear stochastic system is characterized by typical time
scales such as the correlation time of the fast random forces (ff),
$\t_{\text{ff}}$, relaxation times $\t$ of the deterministic part of the
dynamics, the period $T$ of the driving force and the times
$\t_{\text{ms}}$ of typical sojourn within the different metastable
states (ms). In this work we will assume that the correlation
times of the fluctuating forces are much shorter than all other time
scales such that a Markovian description of the dynamics is
appropriate. Hence, we model the fluctuating forces by white noise
($\t_{\text{ff}}=0$)
which moreover will be assumed to be Gaussian and weak. As a
consequence of these assumptions the characteristic sojourn times of the
metastable states are finite but much larger than any of the
deterministic characteristic times ($\t_{\text{ms}} \gg \t$) \cite{HTB}.
This time scale separation implies that the transitions between
the metastable states constitute a discrete Markovian process which
will be investigated in more detail in the present work. We will
demonstrate
that this  discrete process forms the backbone of the original
continuous process on time scales that are much larger than the
deterministic relaxation times $\t$ .

Finally, the magnitude of the driving period $T$ in relation to the
deterministic time scales $\t$ has a decisive influence on the system's
dynamics. In the so-called semiadiabatic limit \cite{T99} the driving
period is large compared to typical deterministic relaxation
times independently of how large the driving period is compared with
the typical sojourn times. Then the time-dependent transition rates
are given by the frozen rates, i.e.\ their time dependence only results
from the slow change of those system parameters that are varied by the
driving process \cite{TL}. Within this framework stochastic resonance
\cite{TMSHL} and the dynamics of neuron models \cite{STH_2004} have
successfully been described.

Outside the regime of the so called semiadiabatic limit the escape rates no
longer instantly follow  but rather lack
behind the periodic driving \cite{STH_2005}. In the present paper we
investigate this regime of intermediate to fast driving in more
detail and present effective methods to characterize the large time
behavior of periodically driven Fokker-Planck processes with
metastable states.

Previous works on periodically driven processes with metastable states
most often have been focussed on particular aspects such as on the
dependence of the average life time of a metastable state \cite{DG,PH1994},
of the exponentially leading part of escape rates within linear
response theory \cite{LMD}, or
on rates in the weak noise limit \cite{LRH,Lehmann2003}.

We close this Introduction with a short outline of the paper.
In Section~\ref{dd} we introduce some important concepts of the
deterministic dynamics of a periodically driven system with coexisting
attractors. In Section~\ref{cpdf} two alternative formulations of the
conditional probability density function are presented for events that
are separated by a time that is much larger than the characteristic
deterministic time $\t$. The first form originates from
the Floquet representation of the
conditional probability density of a periodically driven Markov
process \cite{JH90,JH} while the second expression explicitly refers to the
dynamics of the metastable states. This second expression in particular
contains quantities that
characterize specific probability densities for each metastable state
as well as localizing functions that allocate probabilities to
the metastable states given the
state of the full continuous system. In Section~\ref{lf} we find
equations of motion both for these metastable state specific
probability densities and the localizing functions by
comparing the two
formulations of the conditional probability density at large times.  In
Section~\ref{pdBbo} the theory is exemplified and numerically tested
for a bistable Brownian oscillator. Section~\ref{Con} closes with a
summary.

\section{Characterization of the deterministic dynamics}\label{dd}
In the deterministic limit the considered system is described by the
motion of a state $\bx$ in a $d$ dimensional state space $\S$ governed
by a set
of $d$ coupled differential equations
\be
\dot{\bx} = \bff(\bx,t)\:,
\ee{deq}
where the vector field $\bff(\bx,t)$ periodically depends  on time with
period $T$, i.e.\
$\bff(\bx,t+T) =\bff(\bx,t)$.
We denote the trajectory emanating at the time $s$ from the point $\by$  by
$\bX(t|\by,s)$ and
assume that in the asymptotic limit of large times the motion is
bounded and characterized by a set of $n\geq 2$ different attractors
$\mathcal{A}_\a(t) \subset \S$,
$\a = 1 \ldots n$, such that each trajectory approaches either of the
attractors depending on its initial state and starting time, i.e.\
$\bX(t|\by,s) \to \bx \in \mathcal{A}_\a(t)$ for $t-s$ sufficiently
large. This relaxation process happens
on  a characteristic deterministic time scale of the considered system.
The attractors periodically depend on time, i.e.\
\be
\mathcal{A}_\a(t+T) = \mathcal{A}_\a(t)\: .
\ee{At}
To each  attractor a domain of attraction $\mathcal{D}_\a(s)$
exists
that  consists of all states $\by$ at time $s$ from which the $\a^{\text{th}}$
attractor is reached.
It is formally defined as $\mathcal{D}_\a(s)
= \left \{\by| \bX(t|\by,s) \in \mathcal{A}_\a(t) \;\text{for}\; t-s
  \to \infty\right \}$.
At each fixed time the domains of attraction
form a
partition of the state space into disjoint subsets, which in general
periodically depend on time
\be
\mathcal{D}_\a(t+T) = \mathcal{D}_\a(t)\: .
\ee{Dt}

\section{Conditional probability density of time-periodic Fokker-Planck
processes with metastable  states}\label{cpdf}
\subsection{Floquet representation}\label{Floq}
In many cases the description of a system in terms of deterministic
equations of motion
is sufficient in order to determine the typical behavior of the system
with sufficient accuracy. However, the presence of weak random perturbations,
which often can be modeled by Gaussian white noise, causes different
effects depending on the considered time scales: On characteristic
time scales of the deterministic motion only insignificant deviations
from the deterministic motion typically occur; those trajectories
that start close to the boundaries of the domains of attraction though
are exceptional because they may
be influenced even by small noise, cross the border of the
deterministic domain of attraction and, in this way, come close to a ``wrong''
attractor with finite probability;  all other trajectories are
markedly influenced  on much
longer time scales only on which transitions between the
deterministic, locally
stable states become likely. Hence, these states lose their
stability. Nevertheless, for sufficiently weak noise the system is
found  most of the time close to
one of the formerly stable states. Transitions between these states do
occur with certainty even though this happens rarely. Therefore such
states can be considered as {\it metastable}.

Under the influence of Gaussian white noise the
deterministic dynamical system (\ref{deq})
becomes a Markov process that is characterized by a Fokker-Planck operator
of the following form \cite{HT,R}
\be
L(t) = -\sum_i^d \frac{\partial}{\partial x_i} K_i(\bx,t) + \sum_{i,j}^d
  \frac{\partial^2}{\partial x_i \partial x_j} D_{i,j}(\bx,t)\: .
\ee{FPO}
We here will restrict ourselves to periodically driven processes
where the drift $K_i(\bx,t)$ and possibly also the diffusion $D_{i,j}(\bx,t)$
periodically depend on time with a common period $T$. Hence, $L(t+T)
=L(t)$. 
The time evolution of the system's probability density function (pdf)
$\r(\bx,t)$ is governed by the Fokker-Planck equation
\be
\frac{\partial}{\partial t} \r(\bx,t) = L(t) \r(\bx,t)\: .
\ee{FPE}
In the deterministic limit the diffusion matrix vanishes and
the drift $K_i(\bx,t)$ approaches the deterministic
drift $f_i(\bx,t)$ having the properties discussed in Section~\ref{I}.

A particular solution of the Fokker-Planck equation is the conditional
pdf
 $\r(\bx,t|\by,s)$ to find the process at the state
$\bx$ at time $t$ under the condition that it was at the state $\by$
at time $s$.
It can formally be expressed
in terms of the Floquet representation in the following way \cite{JH90,JH,J,SR,T99}
\be
\r(\bx,t|\by,s) = \sum_i e^{\m_i (t-s)} \ps_i(\bx,t) \f_i(\by,s) \: ,
\ee{FR}
where $\ps_i(\bx,t)$ and $\f_i(\by,s)$ are Floquet eigenfunctions and $\m_i$ are the
corresponding Floquet exponents. They satisfy pairs of mutually adjoint Floquet
equations reading
\begin{equation}
\bsp
\frac{\partial}{\partial t} \ps_i (\bx,t) &= L(t) \ps_i (\bx,t) - \m_i
  \ps_i (\bx,t)\: ,\\
-\frac{\partial}{\partial t} \f_i (\bx,t) &= L^+(t) \f_i (\bx,t) - \m_i
  \f_i (\bx,t)\: ,
\label{FEf}
\end{split}
\end{equation}
with natural boundary conditions with respect to the state variable $\bx$.
Moreover, both types of eigenfunctions are periodic in time
\be
\begin{split}
\psi_{i}(\bx,t+T) &= \psi_{i}(\bx,t)\: ,  \\
\f_{i}(\bx,t+T) &= \f_{i}(\bx,t)\: .
\end{split}
\ee{fpsiT}
The Floquet functions $\ps_i(\bx,t)$ and $\f_j(\bx,t)$ are mutually
orthogonal for eigenvalues $\m_i \neq \m_j$ and can be normalized such
that
\be
\int d\bx \: \f_j(\bx,t) \ps_i(\bx,t) = \d_{i,j}\: ,
\ee{BIO}
where $\d_{i,j}$ denotes the Kronecker symbol.
The Floquet exponents $\m_{j}$ have real parts that are negative or at most
zero.

The representation of the conditional probability in terms of the
Floquet functions further requires that these functions form a
complete set in the sense that
\be
\sum_i \ps_i(\bx,t) \f_i(\by,t) = \d(\bx-\by)\: ,
\ee{CPR}
where $\d(\bx)$ denotes the Dirac $\delta$ function.
We note that equations 
(\ref{FEf}), (\ref{BIO}) and (\ref{CPR}) do not uniquely
determine
the Floquet functions
because gauge transformations of the form
\begin{equation}
\begin{split}
\bar{\psi}_j(\bx,t) &= g_j(t) \psi_j(\bx,t)\:, \\
\bar{\f}_j(\bx,t) &= g_j^{-1}(t) \f_j(\bx,t)\:,\\
\bar{\m}_j &= \m_j + \frac{2 \p i}{T} n_j, \quad n_j \in \mathbb{Z}
\end{split}
\ee{ga}
with gauge factors
\be
g_j(t) = c_j e^{2 \p i n_j t/T}\:, \quad c_j \in \mathbb{C},\; c_j \neq 0
\ee{gj}
generate new
Floquet eigenfunctions, cf. Ref.~\cite{T_2000}. Here $\mathbb{Z}$ and
$\mathbb{C}$ denote the sets of integer and complex numbers,
respectively and $i$ the imaginary unit.

For the sake of definiteness we assume that the gauge chosen for
the Floquet representation of the conditional pdf
(\ref{FR}) is such that the Floquet
exponents assume their smallest possible absolute values.
The Floquet spectrum consisting of these Floquet exponents then
contains the value $\m_0=0$. We assume that this Floquet exponent is
not degenerate \cite{nondeg} if the diffusion matrix is different from
zero.
The corresponding eigenfunction of
$L^{+}(t)$ is constant with respect to $\bx$ and $t$ and can be
chosen as
$\f_0(\bx,t) =1$; the eigenfunction  $\ps_0(\bx,t)$ of $L(t)$ is a
non-negative and normalized function giving the
uniquely defined asymptotic pdf. Hence, it is the unique
solution of the Fokker-Planck equation (\ref{FPE}) that is approached
at time $t$ from any initial state in the remote past at $s \to -
\infty$. As a Floquet
eigenfunction it is periodic in $t$. The normalization
\be
\i_{\S} d\bx\: \ps_{0}(\bx,t)=1
\ee{Nps0}
follows from
eq.~(\ref{BIO}) together with the fact that $\f_{0}(\bx,t) = 1$.

For vanishing noise, the diffusion matrix $D_{i,j}(\bx,t)$ vanishes
and the backward operator becomes a first order
partial differential operator
$L^+_0(t) =  \sum_i f_i(\bx,t) \partial/\partial x_i$
with $f_i(\bx,t)$ being the components of the
deterministic vector field $\bff(\bx,t)$ governing the deterministic
motion, eq.~(\ref{deq}). For a
dynamical system with $n$ coexisting attractors the characteristic
functions of the domains of attraction represent $n$ independent
periodic solutions of the backward equation
$-\partial\f_0/\partial t = L_0^+(t) \f_0 $.
Each of the solutions is unity on one
of the domains of attraction and zero outside. All other periodic
solutions are linear combinations of these characteristic functions.
That means that a deterministic system with $n$ locally stable states
possesses an $n$-fold degenerate Floquet eigenvalue $\m_0 =0$.
As discussed above, in the presence of
noise, the formerly locally stable states become metastable. The
$n$-fold degeneracy of $\m_0 =0$ is
lifted, but at sufficiently weak noise there remains a group of $n$ Floquet
exponents  one of which is exactly zero and the others aquire a small
negative real part. We call them the {\it  slow} Floquet exponents.
For sufficiently small noise this group of slow
Floquet exponents stays well separated from all other Floquet exponents.
For large time lags, the slow Floquet exponents and the
corresponding Floquet eigenfunctions completely determine the
conditional pdf which becomes
\be
\begin{split}
\r(\bx,t|\by,s)& = \sum_{i =0}^{n-1} e^{\m_i(t-s)} \ps_i(\bx,t) \f_i(\by,s)\\
&\qquad \qquad \text{for} \; t-s \gg \t\:,
\end{split}
\ee{rlt}
where the sum only runs over the group of $n$ slow Floquet exponents
i.e.\ over those exponents with the
smallest absolute values.
All other Floquet exponents are determined by the deterministic time
scales all of which are much shorter than those given by the slow
Floquet exponents. Here $\t$ denotes the slowest deterministic time scale.

\subsection{Alternative representation of the conditional probability at large times}\label{acpdf}
In the presence of metastable states the process of moving
from
a state $\by$ at time $s$ to a state $\bx$ at a much later time $t$
may be subdivided into three consecutive steps that correspond to three
contributions to the conditional
probability $\r(\bx,t|\by,s)$:
Within the typical
relaxation time $\t$, compared to which the considered time span $t-s$
is supposed to be very large, the initial state $\by$ will be allocated
to either of the metastable states $\b$ with a
probability $\c_\b(\by,s)$;
within the remaining time $t-s -\t \approx t-s$ the process may visit several
other metastable states and will be found in the
state $\a$ at the final time $t$ with a probability $p(\a,t|\b,s)$.
Given the final discrete state $\a$, the actual continuous states are
distributed
with a pdf $\r(\bx,t|\a)$. For sufficiently small
noise the times within which the first and the last steps are performed
are negligibly short compared to the total time $t-s$. Therefore, the
initial allocation
to a metastable state $\a$  and the final allocation to a continuous
state $\bx$ can be considered as instantaneous events. Moreover, all three
steps are independent of each other and therefore the
conditional probability $\r(\bx,t|\by,s)$ results as
\be
\r(\bx,t|\by,s) = \sum_{\a,\b} \r(\bx,t|\a) p(\a,t|\b,s)
\c_\b(\by,s)\: .
\ee{rpc}
This particular form of the conditional pdf was
derived in the semiadiabatic limit \cite{TL} which is definded by the regime for which the driving is slow compared
to the characteristic local relaxation times but not necessarily slow
compared to the typical transition times between metastable states
\cite{T99}. We claim that this particular form of the conditional
pdf remains to hold true also beyond the
semiadiabatic limit, i.e. in
situations when the driving period is comparable or even faster than the local
relaxation times. The rare occurrence of the transitions between the
metastable states is the only condition required for eq.~(\ref{rpc})
to hold.
It implies the separation of the times needed to perform the first and
the third step compared to the much larger time of the second step and
justifies the independence of these three steps and their respective
contributions to the
conditional probability.
Below, we will infer the main properties of these three sets of
functions $\r(\bx,t|\a)$, $\c_\a(\bx,t)$ and $p(\a,t|\b,s)$
from  their according definitions. \\
(i) Each {\it localizing function} $\c_\a(\bx,t)$ assumes an almost constant
value very close to unity
within the domain of attraction $\mathcal{D}_\a(t)$ and vanishes
outside.
Close to the border of $\mathcal{D}_\a(t)$, the localizing function
 $\c_\a(\bx,t)$
smoothly interpolates between these two values. At each point $\bx$ all
$n$ functions $\c_\a(\bx,t)$ exactly add up to unity:
\be
\sum_\a \c_\a(\bx,t) = 1\: .
\ee{cN}
(ii) Each {\it $\a$-specific pdf} $\r(\bx,t|\a)$ is a strongly
peaked function of $\bx$ about the corresponding attractor
$\mathcal{A}_\a(t)$ and rapidly decays away from the attractor. As
pdf it is
normalized to unity
\be
\int_{\S} d\bx\: \r(\bx,t|\a) = 1 \: ,
\ee{rN}
where the integration extends over the full state space $\S$.
Within the respective domains of attraction $\mathcal{D}_\a(t)$ the
$\a$-specific pdf
almost coincides with the asymptotic
pdf $\ps_{0}(\bx,t)$ up to a normalizing factor.

Property (i) of the localizing function allows one to determine the
probability $p_\a(t)$ of finding the metastable state $\a$
realized at time $t$ for a given pdf $\r(\bx,t)$ in
the following way
\be
p_\a(t) = \int_{\S} d\bx\: \c_\a(\bx,t) \r(\bx,t)\: .
\ee{pa}
On the other hand, one can assign to a given set of probabilities
$p_\a(t)$ a pdf $\r_p(\bx,t)$ by decorating the
metastable states $\a$ with the $\a$-specific pdfs
yielding
\be
\r_p(\bx,t) = \sum_\a \r(\bx,t|\a) p_\a(t)\: .
\ee{rp}
In order that eqs.~(\ref{pa}) and (\ref{rp}) are compatible with each
other, i.e.\ that eq.~(\ref{pa}) reproduces the prescribed probabilities
$p_\a(t)$ for $\r(t) = \r_{p}(t)$, the localizing functions and the
$\a$-specific pdfs must form a biorthonormal set of functions, i.e.\
\be
\int_{\S} d\bx \: \c_\a(\bx,t) \r(\bx,t|\b) = \d_{\a,\b}\: .
\ee{cr}
For a Fokker-Planck process the time evolution of a pdf $\r(\bx,t)$
is determined by the conditional pdf according to
\be
\r(\bx,t) = \int_{\S} d\by\:\r(\bx,t|\by,s) \r(\by,s)\: .
\ee{rrr}
For large time lags $t\!-\!s$ the conditional pdf can be written
as in eq.~(\ref{rpc}). Using eqs.~(\ref{rpc}), (\ref{pa}) and
(\ref{cr}) one obtains from
eq.~(\ref{rrr}) for the propagation of the
probabilities $p_\a(t)$
\be
p_\a(t) = \sum_{\a,\b} p(\a,t|\b,s) p_\b(s)\:.
\ee{ppp}
This relation expresses the occupation probabilities of the metastable
states at a time $t$ in terms of the corresponding probabilities
at an earlier time $s$. Eq.~(\ref{ppp})
hence confirms the interpretation of $p(\a,t|\b,s)$ as the
conditional probability of the coarse grained process of the
metastable, discrete states $\a = 1 \ldots n$.

In order to derive an equation of motion for the probabilities $p_\a(t)$ one
differentiates both sides of eq.~(\ref{pa}) with respect to time, uses
the Fokker-Planck equation (\ref{FPE}), and expresses the pdf by
means of eq.~(\ref{cr}) in terms of the probabilities $p_\b(t)$. In
this way one obtains
\be
\begin{split}
\dot{p}_\a(t)&= \int_{\S} d\bx \: \big \{ \frac{\partial
 \c_\a(\bx,t)}{\partial t} \r(\bx,t)  \\ &\quad + \c_\a(\bx,t) L(t) \r(\bx,t)
\big \} \\
&=\sum_\b k_{\a,\b}(t) p_\b(t)\:,
\end{split}
\ee{me}
where the time dependent rates $k_{\a,\b}(t)$ are defined as
\be
\begin{split}
k_{\a,\b}(t)&= \int_{\S} d\bx\: \frac{\partial\c_\a(\bx,t)}{\partial t}
\r(\bx,t|\b)\\
&\quad + \int_{\S} d\bx\: \c_\a(\bx,t) L(t) \r(\bx,t|\b)\:.
\end{split}
\ee{k}
Eq.~(\ref{cN}) implies that the sum over the first index of the rates
vanishes, i.e.\ $\sum_\a k_{\a,\b}(t) =0$. Therefore, eq.~(\ref{me})
can be brought into the familiar form of a master equation \cite{vK}
\be
\dot{p}_\a(t) = \sum_{\b \neq \a} k_{\a,\b}(t) p_\b(t) - \sum_{\b \neq
  \a}   k_{\b,\a}(t) p_\a(t)\:.
\ee{fme}
We expect that for sufficiently low noise the quantities
$k_{\a,\b}(t)$ do not become negative for $\a \neq \b$ and therefore
represent proper rates. A formal proof of the positivity
though is not available. Negative values of $k_{\a,\b}(t)$ though
would indicate a breakdown
of the basic assumption that the long time behavior of the process is
described by a rate process.
\section{Localizing functions,  $\a$-specific pdfs
  and transition rates}\label{lf}
Comparing the two expressions (\ref{rlt}) and (\ref{rpc}) one finds
that the $\a$-specific pdfs $\r(\bx,t|\a)$ can be
expressed as linear combinations of the
first $n$ Floquet eigenfunctions $\ps_i(\bx,t)$ and the localizing
functions $\c_\a(\bx,t)$ can be written in terms of $\f_i(\bx,t)$. This leads to the
linear relations
\begin{align}
\label{rcp}
\r(\bx,t|\a)&= \sum_{i=0}^{n-1} C_{i,\a}(t) \ps_i(\bx,t)\:, \\
\c_\a(\bx,t)&= \sum_{i=0}^{n-1} D_{\a,i}(t) \f_i(\bx,t)\:,
\label{cdf}
\end{align}
where $C_{i,\a}(t)$ and $D_{\a,i}(t)$ are yet undetermined, time
dependent coefficients.
The orthogonality relations (\ref{BIO}), (\ref{cr}) and the
linear independence of the first $n$ Floquet eigenfunctions
imply the following orthogonality relations of
the coefficients  $C_{i,\a}(t)$
and $D_{i,\a}(t)$:
\be
\begin{split}
\sum_i D_{\a,i}(t) C_{i,\b}(t) &= \d_{\a,\b}\:,\\
\sum_{\a} C_{i,\a}(t) D_{\a,j}(t) &= \d_{i,j}\:.
\end{split}
\ee{CD}
For $i=0$ the normalization of the Floquet function $\ps_{0}(\bx,t)$,
see eq.~(\ref{Nps0}),
and of the
$\a$-specific pdfs $\r(\bx,t|\a)$, see eq.~(\ref{rN}),  leads to
\be
C_{0,\a}(t) = 1.
\ee{C01}
Next we derive sets of coupled equations of motion for the
localizing functions and the $\a$-specific pdfs.

\subsection{Transition rates}\label{tr}
Using the Floquet representation of the $\a$-specific pdfs and
localizing functions, (\ref{rcp}) and (\ref{cdf}), in combination with the
Floquet equations (\ref{FEf})
we obtain for the rates from eq.~(\ref{k})
\be
\begin{split}
k_{\a,\b}(t)& = \sum_i \left ( \dot{D}_{\a,i}(t) C_{i,\b}(t) +
  D_{\a,i} \:\m_i\: C_{i,\b}(t) \right ) \\
&= \sum_i \left ( \dot{D}_{\a,i}(t) D^{-1}_{\b,i}(t) +
  D_{\a,i} \:\m_i\;
  D^{-1}_{\b,i}(t) \right )\:,
\end{split}
\ee{kCD}
where we expressed the
coefficient matrix $C_{i,\b}(t)$ as the inverse of $D_{\b,i}(t)$ by means of
eq.~(\ref{CD}).
Assuming for the moment that the rates $k_{\a,\b}(t)$ were known
we can rewrite
eq.~(\ref{kCD}) as of an equations of motion
for the coefficients  $D_{\a,i}(t)$ and $C_{i,\a}(t)$ reading
\begin{align}
\label{dD}
\dot{D}_{\a,i}(t)& = \sum_{\b} k_{\a,\b}(t) D_{\b,i}(t) -
D_{\a,i}(t) \m_{i}\:,\\
-\dot{C}_{i,\a}(t)& = \sum_{\b} C_{i,\b}(t)k_{\b,\a}(t) -
\m_{i} C_{i,\a}(t)\:.
\label{dC}
\end{align}

It is interesting to note that these are just the Floquet equations of
the master equation (\ref{fme})
and, moreover, that the slow Floquet exponents of the Fokker-Planck
coincide with the
Floquet exponents of the master equation. This is a consequence of the
fact that the master equation specifies the transitions between the metastable
states, and, therefore, represents the backbone of the long time
evolution of the Fokker-Planck process.

With the help of eq.~(\ref{dD}) and the Floquet equations (\ref{FEf})
the following equations of motion for the $\a$-specific
pdfs and the localizing
functions are obtained
\begin{align}\label{er}
\frac{\partial}{\partial t} \r(\bx,t|\a) &= L(t) \r(\bx,t|\a) - \sum_{\b}
k_{\b,\a}(t) \r(\bx,t|\b)\:, \\*[2mm]
-\frac{\partial}{\partial t} \c_\a(\bx,t) &= L^+(t)  \c_\a(\bx,t) -
\sum_{\b}k_{\a,\b}(t)  \c_\b(\bx,t)\:.
\label{ec}
\end{align}
These two sets of equations for the functions $\r(\bx,t|\a)$ and
$\c_{\a}(\bx,t)$  are adjoint to each other such that the
biorthonormality of the $\a$-specific and the localizing functions, see
eq.~(\ref{cr}), continues to
hold for all times once it holds true at a particular instant of time.
Eqs.~(\ref{er}) and (\ref{ec}) represent a central result of this
work.

The set of coupled equations (\ref{er}) can be interpreted as the
motion of $n$ replicas of the original process. Each replica is labeled
by one of the
attractor indices $\a$. The corresponding processes are described by
the Fokker-Planck equation (\ref{FPE}) with  additional
source and sink terms, $\sum_{\b \neq  \a} k_{\b,\a}(t)\r(\bx,t|\a)$
and  $-\sum_{\b\neq \a} k_{\b,\a}(t)\r(\bx,t|\b)$, respectively.
This means that, say, the
$\a$-process dies with probability $\sum_{\b \neq \a}
k_{\b,\a}(t) \r(\bx, t|\b)$ and instantly resurrects
with probability $\sum_{\b \neq \a} k_{\b,\a}(t) \r(\bx,t|\a)$ such
that the total
probability $\i_{\S} d\bx \r(\bx,t|\a)$ of each replica
is conserved for all times. A natural requirement
on a process described by the set of
eqs.~(\ref{er}) is the positivity of the probabilities
$\r(\bx,t|\a)$. For an arbitrary choice of the rates $k_{\a,\b}(t)$ this
property generally will be violated in the course of time. Only for the
correct choice of the transition rates the positivity is guaranteed to
hold. In principle, it is this requirement which determines
the rates $k_{\a,\b}(t)$ on the basis of eq.~(\ref{er}).

In view of the fact that eqs.~(\ref{er}) and (\ref{ec}) are coupled
sets of equations not only for the functions $\r(\bx,t|\a)$  and
$\c_\a(\bx,t)$, respectively, but that in these equations also the
time dependent rates $k_{\a,\b}(t)$ are unknown, it would be very
difficult to solve these equations exactly. Therefore
appropriate approximation schemes have to be devised. This will be
done in the remaining part of this Section.

\subsection{Absorbing boundary approximation: $\a$-specific pdfs}\label{spdf}
Assuming the appropriateness of the rate  description, i.e.\ in
particular the
positivity of $k_{\a,\b}(t)$ for all $\a \neq \b$, one can decompose
the sum on the right hand
side of eq.~(\ref{er}) 
into a sink term $- \sum_{\b \neq \a}
k_{\b,\a}(t) \:\r(\bx,t|\b)$ and a source term $\sum_{\a \neq \b}
k_{\b,\a}(t)\: \r(\bx,t|\a)$. These sink and source terms result from the
diagonal and non-diagonal parts of the rate matrix $(k_{\a,\b}(t))$,
respectively.
The sink terms are linear combinations of the functions $\r(\bx,t|\b)$,
which are strongly
concentrated about the positions of the corresponding attractors
$\mathcal{A}_\b(t)$ with $\b \neq \a$.

We approximate these narrow, even though continuously distributed
sink terms by replacing them
with sharp, absorbing states lying on
the boundaries
$\partial \mathcal{B}_\b(t)$ of domains $\mathcal{B}_\b(t)$. Each
domain $\mathcal{B}_\b(t)$ contains the immediate neighborhood
of
the attractor $\mathcal{A}_\b(t)$ in such a way that the boundary
$\partial
\mathcal{B}_\b(t)$ separates the corresponding attractor from the
remaining state space.
Within this {\it absorbing boundary approximation} we obtain an uncoupled set of equations for
the $\a$-specific pdfs reading
\be
\bsp
\frac{\partial}{\partial t} \bar{\r}(\bx,t|\a) &= L(t)
\bar{\r}(\bx,t|\a) + k_{\a}(t) \bar{\r}(\bx,t|\a)\:,\\
& \qquad \text{for}\;\bx \in
\S_{\a}(t)\:,\\
\bar{\r}(\bx,t|\a) &= 0\:, \quad \text{for all}\; \bx \in
\partial \mathcal{B}_\b(t)\;\text{with}\; \b \neq \a\:,
\end{split}
\ee{ura}
where
\be
k_{\a}(t) \equiv - k_{\a,\a}(t) = \sum_{\b \neq \a} k_{\b,\a}(t)
\ee{ka}
denotes the total decay rate of the state $\a$ which is the sum over
the individual rates from $\a$ to all other states $\b$. The
restricted state space $\S_{\a}(t)$ is obtained from the full state
space $\S$ by excluding the immediate neighborhoods
$\mathcal{B}_{\b}(t)$ of all metastable states $\b$ being different
from $\a$. Hence, it is defined as
\be
\S_{\a}(t) \equiv \S \smallsetminus \cup_{\b \neq \a} \mathcal{B}_{\b}(t)\:.
\ee{Sal}
On this restricted state space the function $\bar{\r}(\bx,t|\a)$ is
expected to represent a valid approximation of the $\a$-specific pdf
$\r(\bx,t|\a)$.

We search
for the periodic solution
of eq.~(\ref{ura}) which can be obtained in the following way. First one
numerically solves the source free problem
\be
\bsp
\frac{\partial}{\partial t} \tilde{\r}(\bx,t|\a) &= L(t)
\tilde{\r}(\bx,t|\a)\:,  \\
\tilde{\r}(\bx,t|\a) &= 0, \quad \text{for all}\; \bx \in
\partial\mathcal{B}_\b(t)\;\text{with}\; \b \neq \a
\end{split}
\ee{tra}
with an initial condition that is positive in a small neighborhood of
the attractor $\mathcal{A}_{\a}(t)$ and vanishes everywhere else.
Because of the absorbing boundary conditions at $\partial
\mathcal{B}_{\b}(t)$, with $\b\neq \a$, the auxiliary function
$\tilde{\r}(\bx,t|\a)$ decays in time, i.e.\
\be
N_\a(t) = \i_{\S_{a}(t)} d\bx \:\tilde{\r}(\bx,t|\a)
\ee{Na}
is a decreasing function of time. Here the integral is extended over
the restricted state space $\S_{\a}(t)$ excluding the domains
$\mathcal{B}_{\b}(t)$, $\b\neq \a$, as defined in eq.~(\ref{Sal}). The
normalized function
\be
\bar{\r}(\bx,t|\a)=\tilde{\r}(\bx,t|\a)/N_\a(t)
\ee{rxa}
 then satisfies the eq.~(\ref{ura}) with the total outgoing rate given by
\be
k_\a(t) = - \frac{\dot{N}_\a(t)}{N_{\a}(t)}\:.
\ee{kaN}
The such constructed solution $\tilde{\r}(\bx,t)/N_{\a}(t)$
approaches a
periodic function in time on the time scale of the deterministic
dynamics, and presents an
approximation to the $\a$-specific function $\r(\bx,t|\a)$.
The other rates $k_{\b,\a}(t)$ leaving the metastable state $\a$
follow from the flux associated with
$\r(\bx,t|\a)$ through the boundaries $\partial \mathcal{B}_\b(t)$
\be
k_{\b,\a}(t) = \i_{\partial \mathcal{B}_{\b}(t)} d\bS \cdot \bj(\bx,t|\a)\:,
\quad \a \neq \b\:,
\ee{kj}
where $d\bS$ denotes the surface element on $\partial \mathcal{B}_{\b}(t)$
pointing towards the metastable state $\mathcal{A}_{\b}(t)$, and
$\bj(\bx,t|\a)$ the probability current carried by the pdf
$\bar{\r}(\bx,t|\a)$. Its components read
\be
\begin{split}
j_{i}(\bx,t|\a) & = K_{i}(\bx,t) \bar{\r}(\bx,t|\a) \\
& \quad - \sum_{l}
\frac{\partial}{\partial x_{l}} D_{i,l}(\bx,t) \bar{\r}(\bx,t|\a)\: .
\end{split}
\ee{j}
This is a generalization of the well known flux-over-po\-pu\-la\-tion
expression for the rate \cite{HTB,fop,K,schmid99}. The stationary flux
carrying pdf of the classical flux-over-population
expression is replaced by the flux carrying time-periodic pdf
$\bar{\r}(\bx,t|\a)$ which is normalized to one, whence also the
population is one.
The decisive difference to the
classical flux-over-population expression lies in the fact that in
eq.~(\ref{kj}) the
flux is determined as the probability flowing per time directly into the final
metastable state, which because of the surrounding absorbing boundary
acts as an outlet,
rather than through a ``saddlepoint'' or ``bottleneck'' on the common
part of the
separatrices $\partial \mathcal{D}_{\a}(t)$ and $\partial
\mathcal{D}_{\b}(t)$ of the initial and the final
metastable state. In the time independent case both expressions
coincide under the
condition that a region containing the final metastable
state and the bottleneck in question is free of sources
\cite{Langer}. In contrast, in the
time-periodic case the probability current contains a periodic
contribution
which in general has a nonuniform phase, i.e.\ the phase depends on the
location $\bx$.
Therefore, the
instantaneous probability flux through the bottleneck in general
differs from the
flux into the outlet. A large portion of probability flowing through
the bottleneck, say within the first half of the period may flow back during
the second half of the period.
Only the time averages over one period of the probabilities flowing
through the bottleneck and into the outlet do coincide.

\subsubsection{$\a$-Floquet functions and rates}\label{Ffr}
The functions $\tilde{\r}(\bx,t|\a)$ which satisfy the
Fokker-Planck equation (\ref{tra}) on the restricted state space
$\S_{\a}(t)$ defined in eq.~(\ref{Sal})
are closely related to the Floquet functions $\ps^{\a}(\bx,t)$ of
the Fokker-Planck operator restricted to $\S_{\a}(t)$ with absorbing
boundaries on the surfaces of the excluded regions
$\mathcal{B}_{\b}(t)$. These $\a$-Floquet functions, as we call them,
are the solutions of the
corresponding Floquet equations which read
\be
\begin{split}
\frac{\partial}{\partial t} \ps_{i}^{\a}(\bx,t)& =
L(t)\ps_{i}^{\a}(\bx,t) - \m_{i}^{\a} \ps_{i}^{\a}(\bx,t)\\
& \qquad
\text{for} \;\bx \in \S_{\a}(t)\:,\;n = 1,2, \ldots\\
\ps_{i}^{\a}(\bx,t)& = 0 \quad \text{for} \; \bx \in \partial
\mathcal{B}_{\b}(t),\; \b \neq \a\:.
\end{split}
\ee{psa}
Because of the absorbing boundaries at all but one metastable states
the Floquet spectrum consisting of the $\a$-Floquet eigenvalues
$\m_{i}^{\a}$ completely lies in the complex half plain with negative
real part. We denote the $\a$-Floquet eigenvalue closest to zero by
$\m_{1}^{\a}$.
The absolute value of the real parts of all other $\a$-Floquet eigenvalues
are much larger, i.e.\ $|\m_{1}^{\a}| \ll |\m_{i}^{\a}| $ for all $i
\neq 1$. In the deterministic limit $\m_{1}^{\a}$ approaches zero,
whereas all other $\a$-Floquet eigenvalues stay finite.

In terms of the $\a$-Floquet eigenfunctions the solution of
eq.~(\ref{tra})  becomes
\be
\tilde{\r}(\bx,t|\a) = \sum_{i=1} c_{i}\: e^{\m_{i}^{\a} t}\:
\ps_{i}^{\a}(\bx,t)\:,
\ee{trpsi}
where $c_{i}$ are constant coefficients whose values depend on the choice of
the initial distribution.
For times which are large on the deterministic time scale, all terms
in the sum become negligibly small apart from the first term
corresponding to $\m_{1}^{\a}$. Hence, we obtain
\be
\tilde{\r}(\bx,t|\a) \propto e^{\m^{\a}_{1} t} \:\ps_{1}^{\a}(\bx,t)\:,
\ee{trps0}
and, by proper normalization
\be
\bar{\r}(\bx,t|\a) = \frac{\ps_{1}^{\a}(\bx,t)}{\i_{\S_{\a}(t)}d\bx
  \:\ps_{1}^{\a}(\bx,t) }\:.
\ee{rps}
With eq.~(\ref{kaN}) the total rate $k_{\a}(t)$ follows as the
negative logarithmic derivative of the normalization $\i_{\S(t)} d\bx\:
\ps_{1}^{\a} (\bx,t)$. It becomes
\be
k_{\a}(t) = -\m_{1}^{\a} + r_{\a}(t)\:,
\ee{kmr}
where
\be
r_{\a}(t) = - \frac{d}{dt} \ln \i_{\S_{\a}(t)} d\bx\: \ps_{1}^{\a}(\bx,t)\:.
\ee{rFp}
The average of $r_{\a}(t)$ over one period vanishes because
$r_{\a}(t)$
is the derivative of a periodic function. Hence, with eq.~(\ref{kmr})
the $\a$-Floquet eigenvalue $\m_{1}^{\a}$ is given by the negative
averaged total rate.

If one performs the time derivative in eq.~(\ref{rFp}) one finds
\be
\begin{split}
r_{\a}(t) &= - \frac{\frac{d}{dt} \i_{\S_{{\a}}(t)} d\bx \:
  \ps_{1}^{\a}(\bx,t)}{\int_{\S_{{\a}}(t)} d\bx \: \ps_{1}^{\a}(\bx,t)} \\
&= -\frac{\i_{\S_{{\a}}(t)}d\bx\: \left [ L(t)\ps_{1}^{\a}(\bx,t) -
  \m_{1}^{\a}\ps_{1}^{\a}(\bx,t) \right ]} {\int_{\S_{\a}(t)} d\bx \:
  \ps_{1}^{\a}(\bx,t)}\\
&=\sum_{\b \neq \a} \frac{\i_{\partial \mathcal{B}_{\b}(t)} \sum_{i,j}
  dS_{i}
  \frac{\partial}{\partial x_{j} }D_{i,j}(\bx,t)
  \ps_{1}^{\a}(\bx,t)}{\i_{\S_{\a}(t)} d\bx \: \ps_{1}^{\a}(\bx,y) }\\
& \quad +
\m_{1}^{\a} \\
&= \sum_{\b \neq \a} k_{\b,\a}(t) + \m_{1}^{\a}  \:.
\end{split}
\ee{rLp}
In the second equality the time derivative was performed. There, the time
dependence of the domain $\S_{\a}(t)$ does not contribute because the
$\a$-Floquet function vanishes on the boundary $ \partial \S_{\a}(t) =\cup_{\b \neq
  \a} \partial \mathcal{B}_{\b}(t)$. The time derivative of
$\ps_{1}^{\a}(\bx,t)$ was expressed by eq.~(\ref{psa}). In the next
step the
integral involving the Fokker-Planck operator was written by means of
Gauss' theorem in terms
of surface integrals over the boundary of $\S_{\a}(t)$.
The terms in the sum on $\b$ are the ratios of the probability fluxes
through the boundaries $\partial \mathcal{B}_{\b}(t)$ carried by the $\a$-Floquet
function $\ps_{1}^{\a}(\bx,t)$, see eq.~(\ref{j}),
and  the corresponding populations
$\i_{\S_{\a}(t)} d\bx \: \ps_{1}^{\a}(\bx,t)$. According to the
eqs.~(\ref{kj}) and (\ref{rps}) the terms in the sum on $\b$ agree
with the individual rates $k_{\b,\a}(t)$.

\subsection{Absorbing boundary approximations: Localizing functions}\label{Lf}
The same type of approximation as for the $\a$-specific pdfs
may also be applied to the equations of
motion for the localizing functions:
By neglecting those terms on the right hand side of eq.~(\ref{ec})
that are proportional
to the rates $k_{\a,\b}(t)$ with $\b\neq\a$ and by introducing absorbing boundary
condititions on the hypersurfaces $\partial \mathcal{B}_\b(t))$, $\b
\neq \a$  we
obtain a set of uncoupled equations for approximate $\a$-localizing
functions $\bar{\c}_{\a}(\bx,t)$ reading
\be
\begin{split}
-\frac{\partial}{\partial t} \bar{\c}_\a (\bx,t) &= L^+(t) \bar{\c}_\a (\bx,t)
+k_{\a}(t) \bar{\c}_{\a}(\bx,t)\:,\\
& \qquad \text {for}\; \bx \in \S_{\a}(t)\:,\\
\bar{\c}_\a(\bx,t) &=0\:, \quad  \text{for all} \;\bx \in \partial
\mathcal{B}_\b(t) \;\text{with}\; \b \neq \a \:.
\end{split}
\ee{c0}
This absorbing boundary approximation is again justified
because the rates $k_{\a,\b}(t)$ are much smaller than the inverse
time scales of the deterministic dynamics which govern the motion
within the domains of attraction.
Moreover it is consistent with the above approximation for the $\a$-specific
pdfs in the sense that the integrals of the products of
the $\a$-specific and the respective localizing
function are independent of time, i.e.\
\be
\frac{d}{dt} \int_{\S_{\a}(t)} d\bx \:\c_{\a}(\bx,t) \r(\bx,t|\a)= 0\:,
\ee{kc}
as follows from eqs.~(\ref{ura}) and (\ref{c0}). Note that the
time dependence of the integration domain $\S_{\a}(t)$ does not
contribute because the integrand vanishes at the boundary.
The biorthonormality of the localizing functions and specific pdfs
cannot be strictly maintained within this approximation. The
deviations though are expected to be exponentially small with respect
to the noise strength because of the small overlap of these
functions for different metastable states.

As in the case of the $\a$-specific functions the total decay rate
$k_{\a}(t)$ need not be known in order to determine the
$\a$-localizing functions. Rather one again may first determine an auxiliary
function $\tilde{\c}_{\a}(\bx,t)$ as the solution of
the source free equation
\be
\begin{split}
-\frac{\partial}{\partial t} \tilde{\c}_\a (\bx,t)& =
L^+(t) \tilde{\c}_\a (\bx,t)\:,\\
\tilde{\c}_{\a}(\bx,t) &= 0, \quad \text{for all}\; \bx \in
\partial\mathcal{B}_\b(t)\;\text{with}\; \b \neq \a\:.
\end{split}
\ee{tc}
Because of the dissipative nature of the backward operator $L^{+}(t)$
it is convenient to integrate this equation backward in time. A
forward integration easily may run into numerical problems because
unavoidable errors would grow exponentially in time.
As an appropriate final condition for $\tilde{\c}_{\a}(\bx,t_{0})$
one may choose a function which
is constant on the domain of attraction $\mathcal{D}_{\a}(t_{0})$ and
zero everywhere else. The solution of this final value problem will
approach a periodic solution on the time scale of the deterministic
dynamics. This asymptotic periodic solution must be normalized at each
instant of
time by the integral of its product with the corresponding
$\a$-specific function to yield the required approximation of
$\c_{\a}(\bx,t)$
\be
\bar{\c}_{\a}(\bx,t) = \frac{\tilde{\c}_{\a}(\bx,t)}{Z_{\a}(t)}\:,
\ee{cZ}
where
\be
Z_{\a}(t) =  \int_{\S_{\a}(t)} d\bx
  \:\tilde{\c}_{\a}(\bx,t) \bar{\r}(\bx,t|\a)\:.
\ee{Zt}
Using the eqs.~(\ref{ura}) and (\ref{tc}) one finds
\be
k_{\a}(t) = \frac{\dot{Z}_{\a}(t)}{Z_{\a}(t)}\:.
\ee{Zk}
This relation confirms that the function given by the eqs.~(\ref{cZ})
and (\ref{Zt})
indeed is a solution of eq.~(\ref{c0}).

\section{Periodically driven Brownian bistable oscillator}\label{pdBbo}
In order to exemplify the theory developed above and to check its
consistency  we consider an overdamped bistable Brownian
oscillator driven by an external force that varies
periodically in time. We choose a bistable quartic potential $V(x,t)$ that
depends periodically on time, see Fig \ref{f1}.
\begin{figure}
\includegraphics[width=8cm]{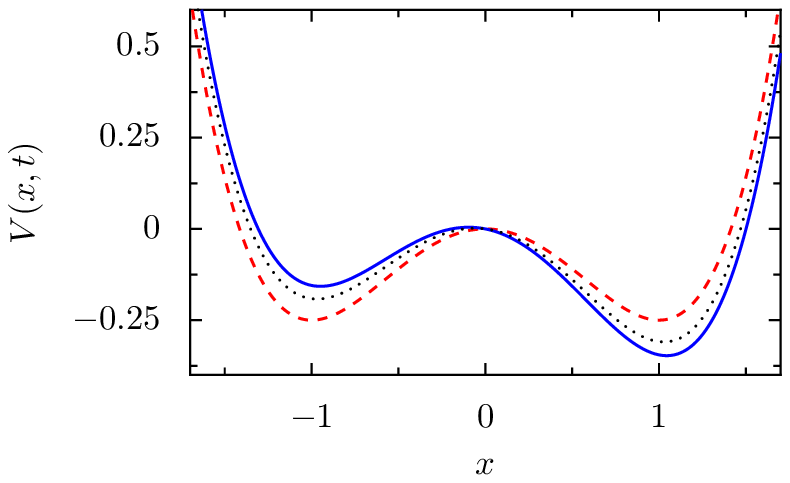}
\caption{ The bistable potential $V(x,t)$,
  eq.~(\ref{Vxt}), is depicted
   as a function of the position $x$ for different times $t=0$ (red, dashed
   line), $t=0.2 T$ (blue, solid line), and $t=0.4 T$ (black, dotted
   line) where $T$ denotes the period of the driving and for the driving
   strength $A=0.1$.}
\label{f1}
\end{figure}
In conveniently chosen dimensionless
variables it reads
\be
V(x,t) = -\frac{1}{2} x^{2} + \frac{1}{4} x^{4} - A x \sin{\O t}\:,
\ee{Vxt}
where $t$ is time and $x$ the position of the Brownian particle. The
strength of the periodic modulation is denoted by $A$ and its
frequency by $\O$.
Depending on the values of $A$ and $\O$ the deterministic overdamped
dynamics in this time dependent potential is either monostable or
bistable as displayed in Fig.~\ref{f2}. In the present context we are
only interested in the bistable region in which the deterministic
dynamics $\dot{x} = -V'(x,t)$ possess two stable limit cycles
$x_{-1}(t)$ and $x_{1}(t)$ and an unstable limit cycle $x_{0}(t)$
forming the separatrix between the two attractors, see Fig~\ref{f3}.
\begin{figure}
\includegraphics[width=8cm]{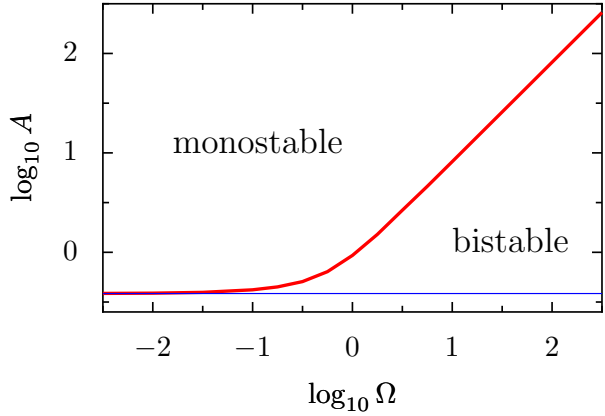}
\caption{The line dividing the $\log_{10} \Omega$ --
  $\log_{10} A$ parameter plane into  an upper monostable and a lower
  bistable region of the deterministic
  dynamics $\dot{x} = - V'(x,t)$ is marked
  by the thick, red solid curve. The blue, thin straight
  line indicates the value of the forcing strength,
  $A^{\text{ad}}= 2/(3 \sqrt{3})$,  below which the potential $V(x,t)$
  has two minima for all times $t$.}
\label{f2}
\end{figure}
\begin{figure}
\includegraphics[width=8cm]{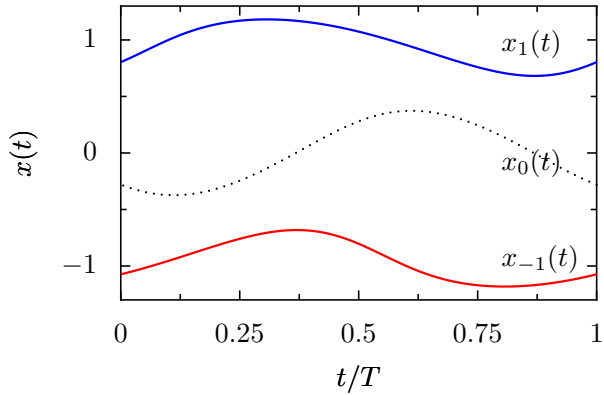}
\caption{ The attractors $x_{-1}(t)$, $x_{1}(t)$ and the separatrix
  $x_{0}(t)$ of the deterministic dynamics $\dot{x} = - V'(x,t)$ for
  the driving strength $A=0.5$ and frequency $\O=1$.}
\label{f3}
\end{figure}
The diffusion matrix $D$ is taken as constant. The Fokker-Planck
operator then becomes
\be
L(t)= \frac{\partial}{\partial x} V'(x,t) + D
\frac{\partial^{2}}{\partial x^{2}}\:,
\ee{LBp}
where $V'(x,t)$ denotes the derivative of the potential with respect
to $x$. The corresponding Fokker-Planck and backward equations were numerically
solved by a collocation method based on a representation of the
solution in terms of Chebishev polynomials of degree 5 \cite{NAG}. For all
calculations a fixed number $N = 1201$ of break-points in the interval
$[-3,3]$ was used. At the ends of the interval reflecting boundary
conditions were imposed.
In the case of the forward equation an accuracy of $10^{-10}$ led to
stable results whereas for the backward equations an accuracy of
$10^{-12}$ turned
out to be necessary in order to avoid numerical artefacts. Throughout
this paper we used a fluctuation strength given by $D=1/40$. At
vanishing driving strength $A=0$ the resulting bistable symmetric
potential  then possesses a barrier height per noise energy of $\D V
/D = \left [V(0,0)-V(1,0) \right ]/D= 10$.
\subsection{Flux-over-population rates}\label{fopr}
We first numerically determined the time dependent solution
$\tilde{\r}(x,t|-1)$ of the
Fokker-Planck equation (\ref{tra}) on the restricted state space
$\S_{-1}(t) = [-3, x_{1}(t)]$ with a reflecting boundary at $x=-3$ and an
absorbing boundary at the
the position of the right attractor $x_{1}(t)$ and
with an initial condition that is sharply
located at the position of the other attractor $x_{-1}(0)$. After a
number $n$ of periods $T=2 \p /\O$ of the driving frequency $\O$ had
elapsed the
remaining population $N_{-1}(t)$ was identified as
\be
N_{-1}(t) = \int_{-3}^{x_{1}(t)} dx \:\tilde{\r}(x,t|-1),
\ee{Nm1}
see also eq.~(\ref{Na}), and the renormalized pdf
\be
\bar{\r}(x,t|-1) =\tilde{\r}(x,t|-1)/N_{-1}(t)\:,
\ee{brm1}
as well as the rate
\be
k_{1,-1}(t) = - \frac{\dot{N}_{-1}(t)}{N_{-1}(t)}
\ee{km1}
were determined. The number $n$ of transient periods was chosen
such that $k_{1,-1}(t)$
remained unchanged upon a further increase of $n$. For different values of
$\O$ appropriate numbers $n$ are collected in Table~\ref{t1}.
 \begin{table}
\caption{Number of transient periods}\label{t1}
 \begin{tabular}{ll}
$\O$ & $n$\\
\hline
1 & 100\\
0.5 & 50\\
0.1 & 10\\
0.01&5\\
0.001&3\\
 \end{tabular}
 \end{table}
In Fig.~\ref{f4} the rates $k_{1,-1}(t)$ are displayed as functions of
time for various driving frequencies. For small frequencies the time
dependent rate approaches its adiabatic form \cite{TL} that is given by the
inverse mean first time that a process needs to move from $x=x_{-1}(t)$ to $x=x_{1}(t)$
in the frozen potential. The rate then reads \cite{HTB}
\be
k^{\text{ad}}_{1,-1}(t) = D \left [\i_{x_{-1}(t)}^{x_{1}(t)} dx \: e^{V(x,t)/D}
  \i_{-3}^{x_{0}(t)} dy \:
  e^{-V(y,t)/D} \right]^{-1}.
\ee{kad}
For larger frequencies the maximal value of the rate shrinks and also
becomes delayed with respect to the driving force. In the limit of
high frequencies it approches the time independent rate
$k^{\text{av}}$ of a Brownian particle moving in the
potential $\overline{V(x,t)} = T^{-1} \i_{0}^{T} dt\: V(x,t)$
averaged over one period of the driving
force.
For the potential given by eq.~(\ref{Vxt}) the average
is symmetric and given by $\overline{V(x,t)} = V(x,0)$.
Hence the rate in the limit of
high driving frequencies
coincides with the value of the adiabatic rate at $t=0$:
\be
k^{\text{av}} = k^{\text{ad}}_{1,-1}(0)\:.
\ee{kav}
Due to the symmetry of the averaged potential, the rate
$k^{\text{av}}$ also describes the opposite transition from the state
$x_{1}(t)$ to $x_{-1}(t)$, whence we skipped the index.

At a fixed frequency the rate $k_{1,-1}(t)$ decreases with decreasing
amplitude $A$ approaching the time independent value $k^{\text{av}}$,
see Fig.~\ref{f5}
\begin{figure}
\includegraphics[width=8cm]{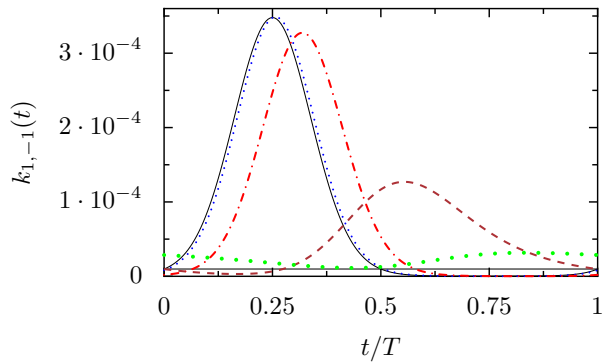}
\caption{ The rate $k_{1,-1}(t)$ following from eq.~(\ref{km1})
  displays a maximum as a function of
  $t/T$ that becomes lower and shifts towards later times within one
  period if the frequency $\O$ increases. For the frequency
  $\O =10^{-3}$ the rate is indistinguishable from the adiabatic rate
  (\ref{kad}) (black, solid line). The other curves display the rates for
  $\O = 10^{-2}$
  (blue, dotted line), $0.1$
  (red, dash-dotted line), $0.5$ (brown, dashed line) and $1$
  (green, thick dots); in the
  asymptotic limit $\O \to \infty$ the constant rate $k^{\text{av}}$
  (thin solid line)
  given by eq.~(\ref{kav}) is approached. In all cases the driving
  strength is $A=0.1$ and the noise strength $D=0.025$.}
\label{f4}
\end{figure}
\begin{figure}
\includegraphics[width=8cm]{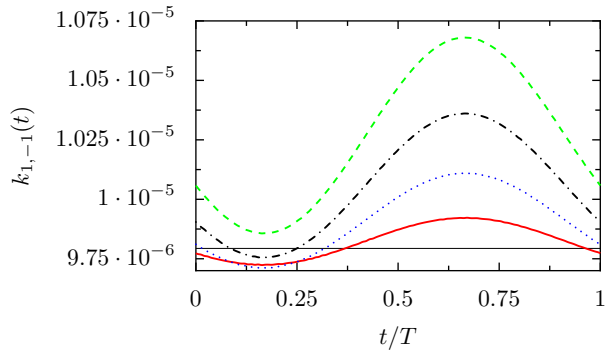}
\caption{ The times at which the rate $k_{1,-1}(t)$
  assumes its extrema do hardly depend on the amplitude $A$.
  The rate is displayed for various values of $A=0.1$ (solid,
red), $0.2$ (dotted, blue), $0.3$ (dashdotted, black), and $0.4$
(dashed, green); in all cases the frequency is
$\O=10$, and the noise $D=0.025$. Note that for the large amplitude
$A=0.4>A^{\text{ad}}$ the deterministic attractors $x_{\pm1}(t)$ are dynamically
stabilized, see also
Fig.~\ref{f2}.}
\label{f5}
\end{figure}

The specific pdf $\bar{\r}(x,t|-1)$ given by
eq.~(\ref{brm1}) represents a
periodic current carrying pdf with an absorbing state
at the attractor $x_{1}(t)$. It possesses a single maximum the location of which closely
follows the deterministic motion of the attractor $x_{-1}(t)$, see
Fig.~\ref{f6}. The pdf is asymmetric about its maximum
with a breathing width that is wider if the maximum is closer to the position of
the separatrix $x_{0}(t)$.

The approximate localizing function $\bar{\c}_{-1}(x,t)$ of the left
metastable state
$x_{-1}(t)$ on the  restricted state space $\S_{-1}(t)$ was obtained from
the solution $\tilde{\c}_{-1}(x,t)$ of the backward equation
(\ref{tc}) with absorbing boundary condition at the right metastable state
$x_{1}(t)$. In order to guarantee for sufficient numerical stability,
the integration of the backward equation has to be performed backward
in time from some $t_{0}$ to times $t<t_{0}$. The final function
$\tilde{\c}_{-1}(x,t_{0})$ was chosen such that it assumes the constant value
$1$ for all $x \in [-3,x_{-1}(t_{0})]$ then decreases  monotonically
and reaches zero at the right metastable state.

After the same
number $n$ of transient periods as for the corresponding characteristic
pdf, see Table~\ref{t1}, the normalization integral (\ref{Zt})
\be
Z_{-1}(t) = \i_{-3}^{x_{1}(t)} dx\: \tilde{\c}_{-1}(x,t) \bar{\r}(x,t|-1)
\ee{Zmt}
was determined.
The rates $k_{1,-1}(t)$ that follow from the logarithmic
derivative of $Z_{-1}(t)$, cf. eq.~(\ref{Zk}), were compared with the
rates obtained from  eq.~(\ref{km1}). They are identical within
numerical accuracy.

Finally, the localizing function $\bar{\c}_{-1}(x,t)$ was determined by
normalizing $\tilde{\c}_{-1}(x,t)$ with $Z_{-1}(t)$. For an example see
Fig.~\ref{f7}. We note that the position where the localizing function
assumes the value $1/2$ coincides with the location of the separatrix at
the respective time.
\begin{figure}
\includegraphics[width=8cm]{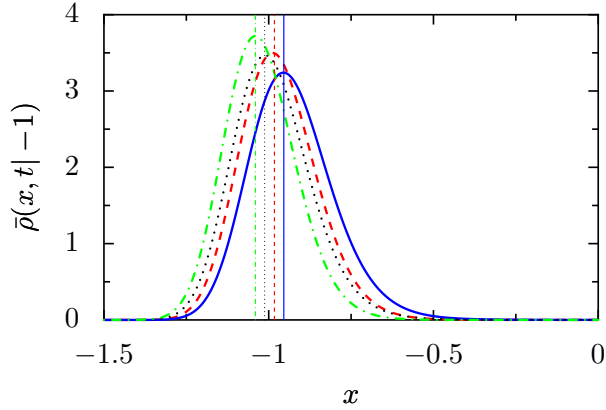}
\caption{ The specific pdf $\bar{\rho}(x,t|-1)$ is
  depicted as a function of the position $x$ for various
  times $t=0.12\:T$ (red, dashed), $0.37\:T$ (blue, solid),
  $0.62\:T$ (black, dotted), and $0.87\:T$ (green, dashed-dotted) for
  the driving
  frequency $\O=1$, driving amplitude $A=0.1$ and noise strength $D =
  0.025$. Outside the displayed interval the specific pdf
  continues to decay. It vanishes at the position of the
  attractor $x_{1}(t)$. The vertical lines indicate the positions of
  the attractor $x_{-1}(t)$ at the respective times. These positions
  almost coincide with the maxima of the specific pdfs
  at the respective times. }
\label{f6}
\end{figure}
\begin{figure}
\includegraphics[width=8cm]{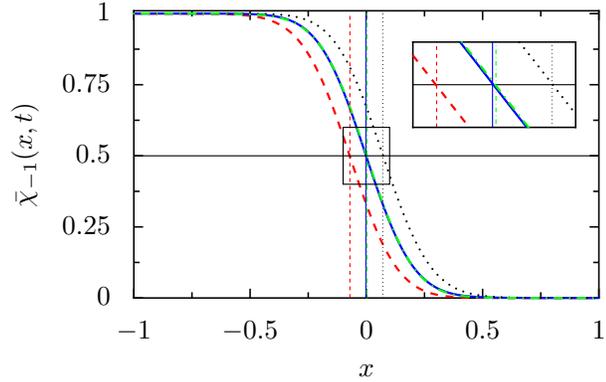}
\caption{ The localizing function $\bar{\c}_{-1}(x,t)$
  interpolates between the values $1$ at the attractor $x_{-1}(t)$ and
  $0$ at $x_{1}(t)$. It is displayed at various instants of time,
  $t=0.12\:T$ (red, dashed), $0.37\:T$ (blue, solid), $0.62\:T$ (black,
  dotted), and $0.87\:T$ (green, dash-dotted). The vertical lines denote the
  positions of the separatrix of the deterministic dynamics at the
  corresponding times, see Fig.~\ref{f3}. In the inset a magnification of the
  center part of the plot marked by a rectangle is depicted. It
  demonstrates that the localizing
  functions very precisely assume the value $1/2$ (horizontal line)
  at the positions of
  the separatrices indicated by the vertical lines.  
}
\label{f7}
\end{figure}
\subsection{Floquet approach}\label{Fa}
Here we construct the specific pdfs and the
localizing functions in terms of Floquet eiegenfunctions on the basis
of the eqs.~(\ref{rcp}) and
(\ref{cdf}). In the present case of two metastable
states these equations simplify to read

\begin{align}\label{Cr}
\r(x,t|\pm 1)&= \ps_{0}(x,t) + C_{\pm 1}(t) \ps_{1}(x,t)\:,\\
\c_{\pm 1}(x,t)&= \frac{C_{\mp 1}(t)}{C_{\mp 1}(t) -C_{\pm 1}(t)}
\nonumber \\
&\quad -\frac{1}{C_{\mp 1}(t)
-C_{\pm 1}(t)}\f_{1}(x,t)\:.\label{Cc}
\end{align}
Here we skipped the first index $i$ of $C_{i,\a}(t)$ since
only the values for $i=1$ are nontrivial in the case of two metastable
states. For $i=0$, $C_{0,\a}(t) =1$ always holds, see eq.~(\ref{C01}).

To further evaluate these equations
(i) the first two Floquet
functions of the forward and the backward equation and (ii) the
coefficients $C_{\pm 1}(t)$ were determined numerically.
The Floquet function $\ps_{0}(x,t)$ belonging to the Floquet
eigenvalue $\m_{0}=0$ is the periodic solution of the
Fokker-Planck equation (\ref{FPE}), (\ref{LBp})
with reflecting boundary conditions at $x=\pm 3$.
As initial condition we chose
\be
\ps_{0}(x,0) = \frac{\exp \left( -  V(x,0)/D\right)}{ \i_{-3}^{3} dx \exp \left(
  -  V(x,0)/D\right)}\:.
\ee{ps0}
The Fokker-Planck equation was numerically solved for $n$ periods of the
driving force.
We designated this number $n$
in such a way that after subsequent $n/10$ periods the $L_{1}$-norm of
the difference of the two solutions was less than $10^{-5}$, i.e.\
\be
||\ps_{0}(x,1.1\:n\: T) -\ps_{0}(x,n\: T)||_{1} \leq 10^{-5}\:,
\ee{L1}
where the $L_{1}$-norm of a function $f(x)$ on the interval $[-3,3]$
is defined by the integral of the its absolute value as
\be
||f(x)||_{1} = \i_{-3}^{3} dx |f(x)| \:.
\ee{L1N}
The numbers $n$ found in this way are collected in Table~\ref{t2} for
different values of the driving frequency.
 \begin{table}
\caption{Number of transient periods needed to reach convergence of
  the Floquet function $\ps_{0}(x,t)$ and Floquet exponent $\m_{1}$}\label{t2}
 \begin{tabular}{lrc}
$\O$ & $n$ & $\mu_{1}$\\
\hline
1 & 10000&-\:4.46 $10^{-5}$\\
0.5 & 2000&- \:9.46 $10^{-5}$\\
0.1 & 1000&-\:1.54 $10^{-4}$\\
0.01&1000&-\:1.58 $10^{-4}$\\
0.001&100&-\:1.58 $10^{-4}$\\
 \end{tabular}
 \end{table}
The Floquet function $\ps_{1}(x,t)$ and the corresponding Floquet
exponent $\m_{1}$ were obtained from the solution of the Fokker-Planck
equation (\ref{FPE}), (\ref{LBp}) with reflecting boundary conditions
at $x=\pm3$ and the initial condition
\be
\tilde{\ps}_{1}(x,0) = \d\big (x-x_{-1}(0) \big )\:.
\ee{ps1}
After a transient period of duration $n\:T$ with $n$ given by
Table~\ref{t1} the
logarithm of the $L_{1}$-norm of the difference between
$\tilde{\ps}_{1}(x,t)$ and $\ps_{0}(x,t)$ was plotted as a function of
time for several periods. Its logarithm $\ln ||\tilde{\ps}_{1}(x,t) -
\ps_{0}(x,t)||_{1}$ is the superposition of a declining
linear  and a periodic function of time with period $T$ of the
driving.
The Floquet exponent $\m_{1}$
can be read off from the inclination of the linear contribution. The results are
presented in Table~\ref{t2}. We note here that the method of the
$\a$-Floquet functions defined on a restricted phase space with an
absorbing state at, say $x_{1}(t)$, see Section~\ref{Ffr},
gave  Floquet exponents
$\m_{1}^{-1}$ which coincide with those based on the full state space
up to 4 or 5 digits. The same agreement was obtained from the time average
of the rates obtained by either of the methods described in the
previous Section~\ref{fopr}.
Once the Floquet exponent $\m_{1}$ is known, the still unnormalized Floquet
eigenfunction is obtained as
\be
\ps_{1}(x,t)= e^{-\m_{1} t} \left (\tilde{\ps}_{1}(x,t) - \ps_{0}(x,t)\
\right)\:.
\ee{ps1n}
The first two Floquet eigenfunctions, which were normalized with
respect to the $L_{1}$-norm, are displayed in Fig.~(\ref{f8}).
\begin{figure}
\includegraphics[width=8cm]{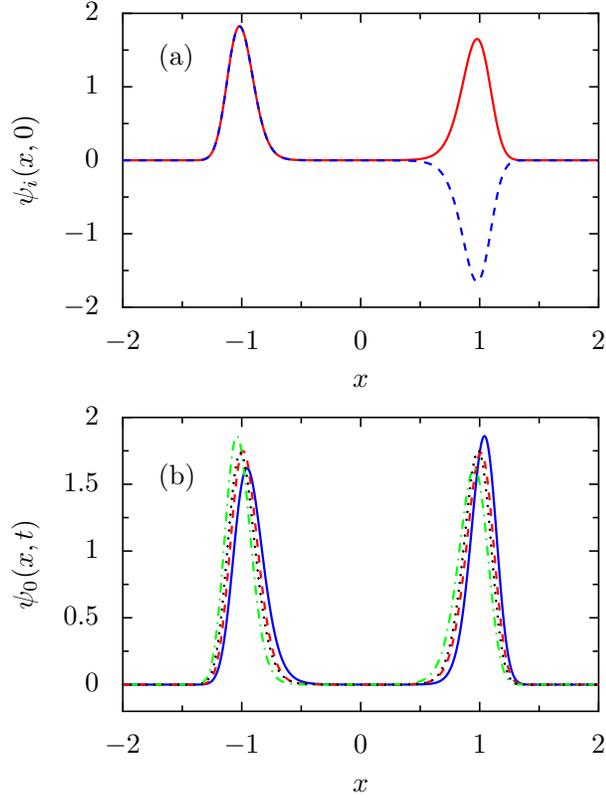}
\caption{ The first two Floquet eigenfunctions
  $\ps_{0}(x,0)$ (red, solid line)
  and
  $\ps_{1}(x,0)$  (blue, dashed line) of the Fokker-Planck operator
  (\ref{LBp}) of a
  driven Brownian oscillator in a bistable potential (\ref{Vxt}) for
  the driving strengths $A=0.1$, driving frequency $\O=1$ and noise
  strength $D=2.5\times 10^{-2}$ at $t=0$ that are displayed in panel (a)
  are strongly localized in the vicinity of the two metastable states
  at $x_{\pm1}(0)$. Both functions are normalized such that their
  $L_{1}$-norms are one, i.e.\ $||\ps_{i}(x,t)||_{1} = \i_{-3}^{3}dx
  |\ps_{i}(x,t)| = 1$. The two functions almost agree with each other up
  to a change in sign close to the unstable point $x_{0}(0)$. In panel
  (b), the time
  dependence is indicated for the asymptotic pdf $\ps_{0}(x,t)$ for
  four different times $0.12 T$ (red, dashed line), $0.37 T$ (blue,
  solid line),  $0.62 T$ (black, dotted line)
  and $0.87T$ (green, dash-dotted line).
}
\label{f8}
\end{figure}
The Floquet eigenfunction of the backward operator
belonging to the Floquet exponent $\m_{0}=0$
is known to be constant, i.e.\ $\f_{0}(x,t) = 1$. In order to determine the
Floquet eigenfunction $\f_{1}(x,t)$ belonging to $\m_{1}$ we solved
the backward equation
\be
-\frac{\partial }{\partial t} \tilde{\f}_{1}(x,t) =
L^{+}(t)\tilde{\f}_{1}(x,t)
\ee{bwe}
with the initial condition
\be
\tilde{\f}_{1}(x,0)= \text{sign}(x) \cdot \left \{
\begin{array}{ll}
-1&|x|\geq 0.1\\
100 \cdot (|x|-0.1)^{2}-1 \quad&|x| \leq 0.1\:.
\end{array}
\right .
\ee{f1i}
After a transient time of duration $nT$ with $n$ given in
Table~\ref{t1} all contributions from higher Floquet functions have
become negligible and $\tilde{\f}_{1}(x,t)$ assumes the form
\be
\tilde{\f}_{1}(x,t) = c_{0} + e^{\m_{1}t} c_{1} \f_{1}(x,t)\:.
\ee{f1t}
Knowing the Floquet exponent $\m_{1}$ we determined the constant $c_{0}$
such that $[\tilde{\f}_{1}(x,t) -c_{0}]\exp(-\m_{1} t)$ becomes a
periodic function of time which is proportional to the sought-after
function $\f_{1}(x,t)$.
The normalization of $\f_{1}(x,t)$ is chosen such that
\be
\i_{-3}^{3} dx \:\f_{1}(x,t) \ps_{1}(x,t) = 1\:.
\ee{fps}
The spatial and temporal dependence of $\f_{1}(x,t)$ is depicted in
Fig.~\ref{f9} for the same parameter values as for the periodic pdf
displayed in Fig.~\ref{f8}.
\begin{figure}
\includegraphics[width=8cm]{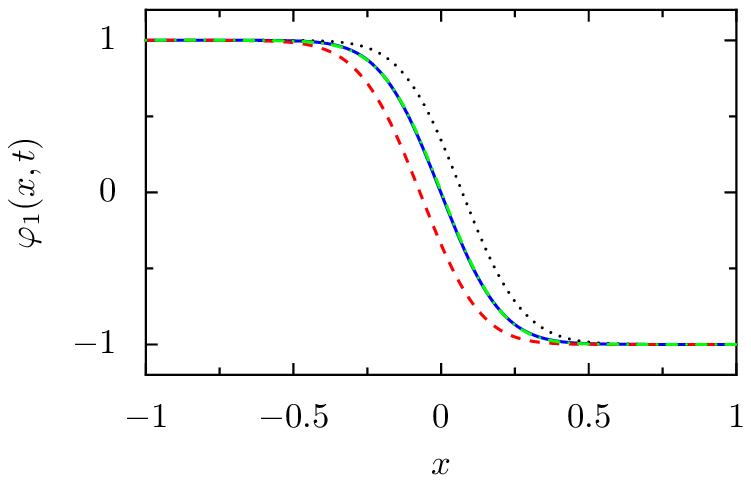}
\caption{The Floquet eigenfunctions $\f_{1}(x,t)$
  of the backward operator for the times $0.12 T$ (red, dashed line),
  $0.37 T$ (blue, solid line), $0.62 T$ (black, dotted line)
  and $0.87T$ (green, dash-dotted line)
  are almost constant apart from a narrow
  region about the unstable fixed point $x_{0}(t)$. The parameters are
  with $A=0.1$, $\O=1$ and $D=2.5\times 10^{-2}$ the same as in Fig.~\ref{f8}.
 }
\label{f9}
\end{figure}

Once the Floquet functions $\ps_{i}(x,t)$ for $i=0,1$ are known the
coefficients $C_{\pm1}(t)$ can be determined from the condition that
the $\a$-specific pdf $\r(x,t|\a)$ is negligibly small in the vicinity
of the other metastable state $x_{\b}(t)$ ($\a \neq \b$). Hence the
intergration on both sides of eq.~(\ref{Cr}) over a small neighborhood
of $x_{\mp}(t)$ gives a negligibly small contribution and thus leads to
the following expression for the coefficients $C_{\pm1}(t)$
\be
C_{\pm1}(t) \approx -\frac{\i_{x_{\mp1}(t)-\e/2}^{x_{\mp1}(t)+\e/2}
  dx\:\ps_{0}(x,t)}{\i_{x_{\mp1}(t)-\e/2}^{x_{\mp1}(t)+\e/2}
  dx\:\ps_{1}(x,t)}\:.
\ee{Cpm}
As an example the coefficient $C_{-1}(t)$ is displayed in
Fig.~\ref{f10} for different values of the driving frequency. The
interval length was chosen as $\e=0.1$.
\begin{figure}
\includegraphics[width=8cm]{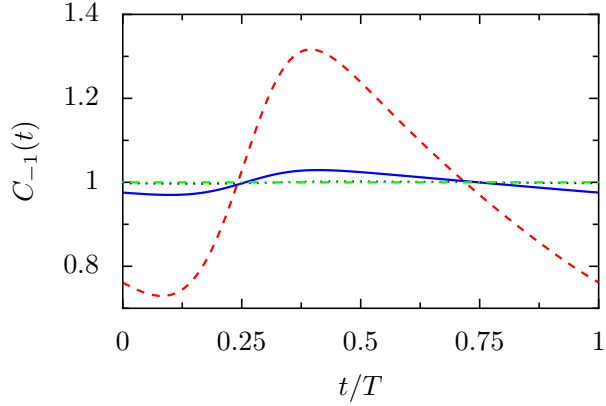}
\caption{ The variability of the coefficient $C_{-1}(t)$
  within one period $T$ of the driving decreases with increasing
  frequency $\O= 10^{-3}$ (red, dashed), $10^{-2}$ (blue, solid),
  $10^{-1}$ (black, dotted) and $1$ (green, dash-dotted). The other
  parameters are with $A=0.1$ and  $D=2.5\times 10^{-2}$ the same as in Fig.~\ref{f8}.
 }
\label{f10}
\end{figure}
Once the first two Floquet eigenfunctions and the coefficients
$C_{\pm1}(t)$ are known, the specific pdfs $\r(x,t|\pm1)$ and the
localizing functions $\c_{\pm1}(x,t)$
can be calculated and compared with the results for
$\bar{\r}(x,t|\pm 1)$ and $\bar{\c}_{\pm1}(x,t)$, respectively, obtained by the
flux-over-population method. We here restrict ourselves to a
comparison for the specific pdf $\r(x,t|-1)$ for fast driving with $\O=1$.
Fig.~\ref{f11} demonstrates the perfect agreement. Only in the immediate
vicinity of the metastable state a difference becomes visible upon
strong magnification.
\begin{figure}
\includegraphics[width=8cm]{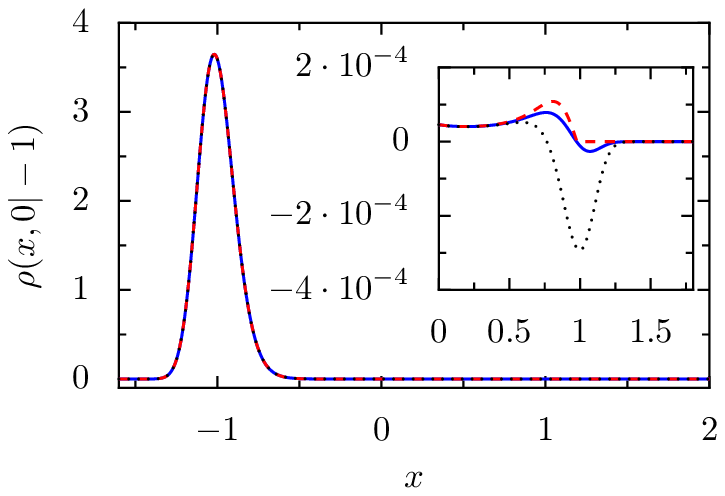}
\caption{ The specific pdf $\r(x,0|-1)$ was determined by
  three different methods: As the flux carrying periodic pdf
  $\bar{\r}(x,t|-1)$ in the presence of a sharp absorbing boundary at
  $x_{1}(t)$ (red, dashed line),  and as a linear combination of the
  first two Floquet eigenfunctions, see eq.~(\ref{Cr}), with coefficients
  either determined by eq.~(\ref{Cpm}) (blue, solid line), or from the
  solutuion of the Floquet problem of the master equation (black,
  dotted line), see the discussion below. Only in the magnification
  displayed in the inset a deviation of the results of these methods
  becomes visible in the vicinity of the metastable state
  $x_{1}(0)\approx 0.98$ where $\bar{\r}(x_{0}(0),0|-1) =0$. We expect that these
  small deviations become even smaller at smaller noise strength.
  }
\label{f11}
\end{figure}

Moreover, from the coefficients $C_{\pm1}(t)$ and the Floquet exponent
$\m_{1}$ the rate $k_{-1,1}(t)$ and $k_{1,-1}(t)$ can be determined
according to eq.~(\ref{dC}) which simplifies for $k_{1,-1}(t)$ in the case of two
metastable states to
\be
k_{1,-1}(t) =\frac{\m_{1} C_{-1}(t) -
  \dot{C}_{-1}(t)}{C_{1}(t)-C_{-1}(t)}\: .
\ee{kpmC}
A comparison of these rates with those obtained by the reactive flux
method is presented in Fig.~\ref{f12} for different values of the
driving frequency. A qualitatively good agreement is obtained for all
frequencies whereby deviations become more visible for higher
frequencies.
\begin{figure}
\includegraphics[width=8cm]{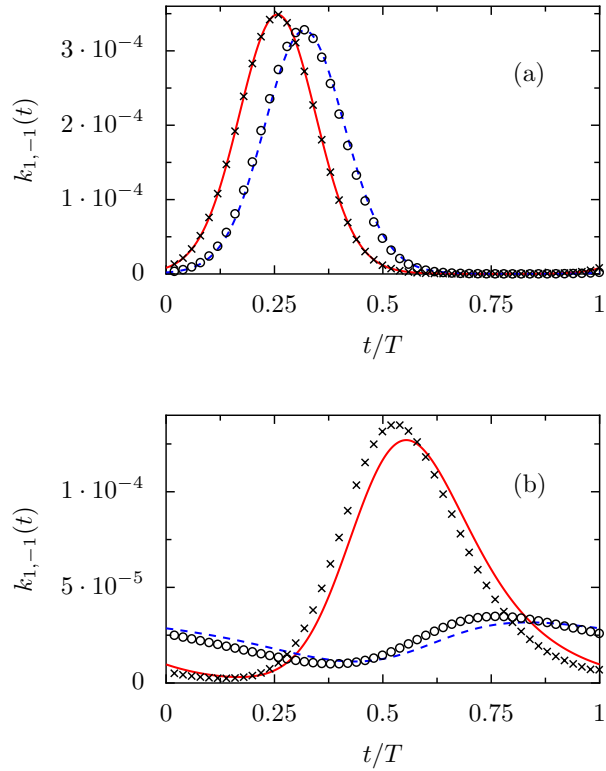}
\caption{A comparison of the flux-over-population
  rates (fop rates) (lines)
 with the Floquet rate expressions (F rates) following from
 eq.~(\ref{kpmC}) (symbols)
 is presented for frequencies $\O=0.01$ (fop rates: red, solid line; F
 rates: crosses) and $\O=0.1$ (fop rates: blue, dashed line; F rates:
 circles) in panel (a), and for $\O=0.5$ (fop rates: red, solid line; F
 rates: crosses) and  $\O=1$ (fop rates: blue, dashed line; F rates:
 circles) in panel (b). The remaining parameters are with $A=0.1$, $D=2.5\times
 10^{-2}$ the same as in Fig. \ref{f8}.
  }
\label{f12}
\end{figure}
\subsection{Decoration}\label{deco}
Finally, we numerically investigated the crucial assumption that
after a sufficiently large transient period the pdf $\r(\bx,t)$
takes the form of eq.~(\ref{rp}), i.e.\ it is determined
by the solutions of the master equation (\ref{fme}), $p_{\a}(t)$,
which are decorated by the $\a$-specific pdfs
$\r(\bx,t|\a)$. As a quantitative measure
of the distance between the numerically exact solution $\r(x,t)$ of
the Fokker-Planck equation (\ref{FPE}), with the Fokker-Planck
operator (\ref{LBp}),
starting at the metastable state
$x_{-1}(0)$, i.e.\ with the initial condition (\ref{ps1}),
and an approximate form $\rho_{\text{a}}(x,t)$ of the pdf we employed the
$L_{1}$-norm (\ref{L1N}) of the difference of these functions.
The assumed asymptotic form
\be
\r_{\text{a}}(x,t) = \r(x,t|1) p_{1}(t) + \r(x,t|-1) p_{-1}(t)
\ee{r}
requires the knowledge of the probabilities $p_{\pm1}(t)$ which was
obtained as the solution of the master equation
\be
\begin{split}
\dot{p}_{1}(t)& = -k_{-1,1}(t) p_{1}(t) + k_{1,-1}(t) p_{-1}(t)\\
\dot{p}_{-1}(t)& = k_{-1,1}(t) p_{1}(t) - k_{1,-1}(t) p_{-1}(t)\\
p_{1}(0)&=0\:, \quad
p_{-1}(0) =1\:,
\end{split}
\ee{mets}
where the flux-over-population expressions were taken for the rates,
see Section \ref{fopr}.
For the specific pdfs we employed three different approximations:
First we used the current carrying pdfs $\bar{\r}(x,t|\pm 1)$ introduced in
Section~\ref{fopr}. These functions were extended onto the full state space
$[-3,3]$ by assigning the value zero beyond their respective domains of
definition, i.e.\ we defined
\be
\begin{split}
\r_{I}(x,t|\!-\!1)& = \left \{
\begin{array}{ll}
\bar{\r}(x,t|\!-\!1) \;\:& \text{for}\; -3\leq x \leq x_{1}(t)\\
0& \text{for}\; x_{1}(t)\leq x \leq 3
\end{array}
\right .\\
\r_{I}(x,t|1)& = \left \{
\begin{array}{ll}
0 & \text{for} \;-3\leq x \leq x_{-1}(t)\\
\bar{\r}(x,t|1) \;\quad&\text{for}\; x_{-1}(t)\leq x \leq 3\:.
 \end{array}
\right .
\end{split}
\ee{rI}

As a second and third approximation, in the followowing referred to as
approximation II
and III, we used the specific pdfs (\ref{Cr}) with the numerically
determined Floquet functions, see Section~\ref{Fa}, and
determined the coefficients $C_{\pm1}(t)$ in two different ways.
The approximation II was
obtained by using eq.~(\ref{Cpm}) for the coefficients $C_{\pm1}(t)$.
The approximation III is based on the fact that these coefficients
obey the Floquet equations (\ref{dC}) of the backward master
equation. We numerically solved these equations under the assumption
that the rates are given by the flux-over-population expressions.
The resulting functions $c_{\pm1}(t)$ then coincide with the
sought-after coefficients $C_{\pm1}(t) = q c_{\pm1}(t)$ up to a common
proportionality constant $q$. Finally this coefficient was determined
such that the distance between the numerical solution of the
Fokker-Planck equation and the approximation III, i.e.\ $||\r(x,t) -
\r_{\text{III}}(x,t)||_{1}$, became minimal at $t=n T$ with $n$ from
Table~\ref{t1}.
The coefficients $C_{\pm1}(t)$ obtained in this way are compared with
those used in the approximation II, see Fig~\ref{f13}.
\begin{figure}
\includegraphics[width=8cm]{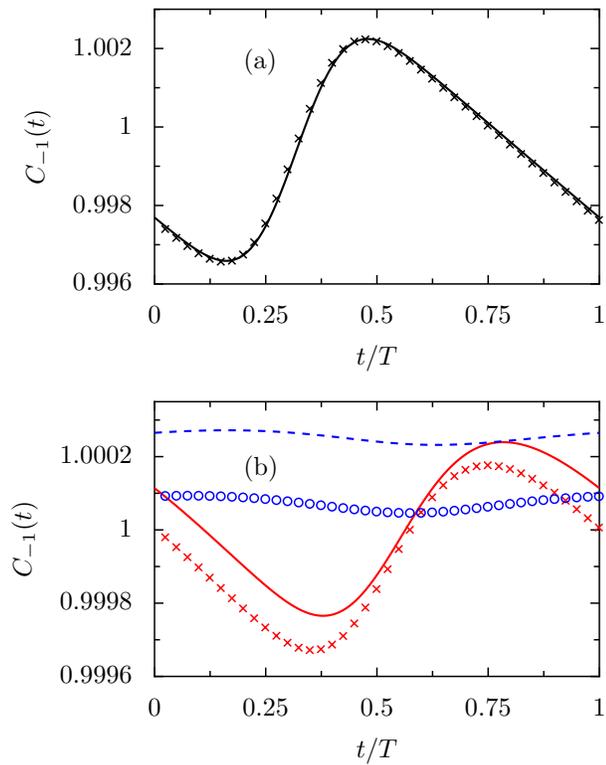}
\caption{The comparison of the approximations II and III
  for the coefficient $C_{-1}(t)$ shows perfect agreement for driving frequencies
  $\O\leq 0.1$, see  panel (a) for $\O=0.1$ (method II: crosses, method
  III: solid line). Relatively small but on the scale of the
  variability apparent deviations between the methods become visible for
  $\O=0.5$ (red, method II: crosses, method III: solid line) and $\O=1$
  (blue, method II: circles, method III: dashed line) in panel
  (b). The remaining
  parameters in both panels are with $A=0.1$, $D=2.5\times
  10^{-2}$ the same as in Fig. \ref{f8}.
  }
\label{f13}
\end{figure}
The relative deviation between the coefficients $C_{\pm}(t)$ resulting
from the approximations II and III were smaller than $5\times 10^{-4}$ in all
investigated cases. Clear deviations are visible only on the scale of
the variability of the coefficients for frequencies $\O>0.1$, see
Fig~\ref{f13}.
\begin{figure}
\includegraphics[width=8cm]{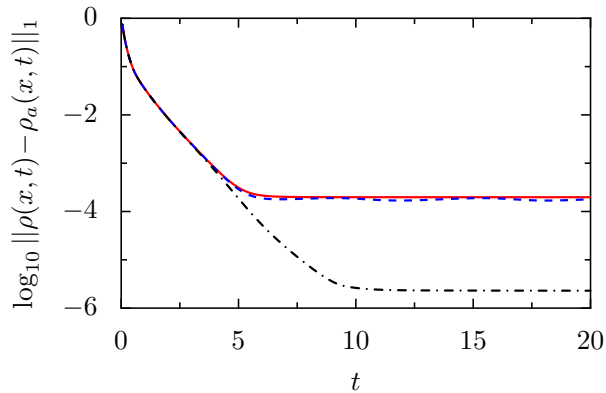}
\caption{ After a short relaxation time, the decadic
  logarithm of the $L_{1}$ distance
  between the numerical solution of the Fokker-Planck equation and
  the proposed asymptotic form (\ref{r}) reveals a perfect
  agreement with $\r_{\text{III}}$ within the expected numerical
  precision of the solution of the Fokker-Planck equation (black,
  dash-dotted line). In the case
  of the first method (red, solid line) which uses the decoration with
  the current
  carrying densities, the absorbing boundary conditions at one of the
  metastable states leads to a larger distance from the asymptotic
  pdf. This also happens with method II (blue, dashed line) which is
  based on the estimate
  (\ref{Cpm}) of the coefficients $C_{\pm 1}(t)$ which lacks a rigorous
  foundation. Yet the observed agreement is very good even for rather
  fast driving with the frequency $\O=1$.
  The remaining parameters are with $A=0.1$, $D=2.5\times
  10^{-2}$ the same as in Fig. \ref{f8}.
  }
\label{f14}
\end{figure}

The distances between the numerically exact solution of the
Fokker-Planck equation and the pdfs obtained from the decoration of
the metastable states according to the three methods described above
are displayed in Fig.~\ref{f14}. In all cases, after a short initial
time,  an exponential relaxation sets in until the pdfs obtained from
method II as well as from the
decoration with the current carrying pdfs   saturate at a distance
of the order of $2\times10^{-4}$. For method III it does so at the
smaller distance of $2\times 10^{{-6}}$. This is a clear indication that
the asymptotic pdf is indeed of the form of eq.~(\ref{r}). This hence
corroborates a basic assumption of our work about the structure of the
pdf at large times.

\section{Summary}
\label{Con}
We investigated the large time stochastic dynamics of  periodically
driven systems with metastable states governed by a Fokker-Planck
equation.
On time scales larger than the typical deterministic
time scale this dynamics can be completely characterized by the
localizing functions, the $\a$-specific pdfs and the conditional
occupation probabilities of the metastable states. The latter are
solutions of a Markovian master equation with time-dependent
rates. These rates can be expressed in terms of the localizing functions
and the $\a$-specific pdfs, see eq.~(\ref{k}).

Using the Floquet
representation of the conditional pdf in the large time limit we
obtained coupled equations of motion for the $\a$-specific densities
and an adjoint set of equations for the localizing functions. Most
interestingly, these equations of motion can be interpreted in the
spirit of Farkas' \cite{fop} and Kramers' \cite{K} idea to construct a
flux carrying stationary solution by imposing convenient sources and
sinks. 
To each $\a$-specific density an $\a$-process can be assigned that
evolves according to the same 
dynamical laws as
the original process with the only difference that it can instantly
be translocated in state space.
These translocations are governed by sinks and sources that cause 
a sudden death of an $\a$-process, say, at
a point $\bx$ and the instant resurrection of the same process at a
different point $\by$ in state space. The sinks are determined by the 
sum of transition
rates out of the metastable state $\a$ multiplied by those $\b$ specific pdfs
corresponding to states that can directly be reached from $\a$.
The source is given by the total rate to leave state $\a$ multiplied
by its specific pdf.
In this way the conservation of probability of each specific
pdf is guaranteed. Due to the resulting
intricate coupling and the dependence on the
unknown rates, an exact solution is difficult to construct and one
must rely on approximate methods to solve this set of equations of motion
for the $\a$-specific pdfs.

An efficient way of
approximation is based on the fact that at weak noise the
$\a$-specific pdfs are expected to be strongly localized in the region
of the according metastable state. This allows one to effectively
decouple the equations for the $\a$-specific pdfs (as well as those for
the localizing functions) and to calculate a current carrying pdf in
the presence of sharply absorbing states. The rates of all transitions
leaving the considered metastable state can then be
calculated by means of a flux-over-population expression \cite{fop,K,schmid99}. In contrast
to the case without time-dependent driving it is important to
calculate the probability flux flowing directly
into the final metastable state.
In the time independent case this flux is the same through all
hypersurfaces in state space separating
the initial from the final metastable state. In the presence of
periodic driving the total flux through a hypersurface in general
depends both on time and on the location of the chosen hypersurface.
The proper rate therefore must
be determined from the probability flux flowing directly into the final
metastable state.

We illustrated our theory with the example of a periodically
driven bistable Brownian oscillator. In contrast to a slowly driven
bistable oscillator, at finite frequencies bistability
extends to larger amplitudes of the driving force.
We found that the flux-over-population method based on the $\a$-specific
pdf with an absorbing boundary at the final metastable state requires
a much lesser computational effort than the direct application of the
Floquet approach. In the former case the solution of the Fokker-Planck
equation with the appropriate boundary conditions converges on the
order of the deterministic time scale, whereas for the second method
the convergence of the Floquet functions is only reached after several
transitions between the metastable states have taken place on average.

We note that based on the absorbing boundary approximation the
transition rates can also be determined by means of numerical
simulations of the Langevin equations of the considered Fokker-Planck
process \cite{TMSHL,STH_2004,STH_2005}.

We finally tested the crucial assumption of our theory saying that the
probability density resulting as the large time solution of the Fokker-Planck
equation
can be represented as the product of the probabilities of the
metastable states decorated by the specific pdfs. The time dependence
of the probabilities of the metastable states was obtained from the
solution of the master equation with the numerically determined
flux-over-population rates.  The specific pdfs obtained by the
absorbing boundary approximation already lead to an excellent
agreement with the numerically exact solution of the Fokker-Planck equation
on time scales larger than a few characteristic deterministic times.
A more elaborate calculation of the specific pdfs in terms of
Floquet eigenfunctions of the Fokker-Planck operator led to a further
improvement of the agreement by two orders of magnitude confirming our
assumption.

\section*{Acknowledgments}
Two of us (P.T. P.H.) like to acknowledge innumerable 
stimulating and provocative scientific discussions with Eli Pollak who
is still at an age well fitted to appreciate   
and to contribute great science.
This work was supported by the DFG via research
center, SFB-486, project A10, via the project no. 1517/26-2, the
Volkswagen Foundation (project I/80424), the German Excellence
Initiative via the \textit {Nanosystems Initiative Munich} (NIM), and
Research Foundation funded by the Korean Government
(MOEHRD), Basic Research Promotion Fund Grant No.
KRF-2005-070-C00065, by the Korea Science and Engineering
Foundation Grant No. F01-2006-000-10194-0, and by
the Deutsche Forschungsgemeinschaft and the Korea Science
and Engineering Foundation in the framework of the
joint KOSEF-DFG grant no. 446 KOR 113/212/0-1.



\begin{thebibliography}{99}
\bibitem{nondeg} The Floquet eigenvalue $\m_{0} =0$ is non-degenerate
  if the diffusion matrix is non-singular.
  There are also processes with singular diffusion
  matrices such as for Fokker-Planck operators of the
  Klein-Kramers type for which   $\m_{0} =0$ is still non-degenerate.
\end{thebibliography}
\end{document}